\documentclass[12pt,preprint]{aastex}

\usepackage{graphicx}
\usepackage{amssymb}
\usepackage{amsmath}

\newcommand{\new}[1]{{\rm #1}}
\defcitealias{Lane09a}{Paper I}
\defcitealias{Lane09b}{Paper II}

\shorttitle{A Merger in 47 Tuc}
\shortauthors{Lane et al.}

\begin{document}

\title{AAOmega Observations of 47 Tucanae:\\Evidence for a Past Merger?}

\author{Richard R. Lane\altaffilmark{1}, Brendon
  J. Brewer\altaffilmark{2}, L\'aszl\'o L.  Kiss \altaffilmark{1,3},
  Geraint F. Lewis\altaffilmark{1}, Rodrigo A.  Ibata\altaffilmark{4},
  Arnaud Siebert\altaffilmark{4}, Timothy R.  Bedding\altaffilmark{1},
  P\'eter Sz\'ekely\altaffilmark{5} and Gyula
  M. Szab\'o\altaffilmark{3}}

\altaffiltext{1}{Sydney Institute for Astronomy, School
  of Physics, A28, The University of Sydney, NSW, Australia 2006}
\altaffiltext{2}{Department of Physics, University of California, Santa
    Barbara, CA, 93106-9530, USA}
\altaffiltext{3}{Konkoly Observatory of the
    Hungarian Academy of Sciences, PO Box 67, H-1525, Budapest,
    Hungary}
\altaffiltext{4}{Observatoire Astronomique, Universite de
    Strasbourg, CNRS, 67000 Strasbourg, France}
\altaffiltext{5}{Department of
    Experimental Physics, University of Szeged, Szeged 6720, Hungary}

\begin{abstract}
\noindent The globular cluster 47 Tucanae is well studied but it has
many characteristics that are unexplained, including a significant
rise in the velocity dispersion profile at large radii, indicating the
exciting possibility of two distinct kinematic populations.  In this
Letter we employ a Bayesian approach to the analysis of the largest
available spectral dataset of 47 Tucanae to determine whether this
apparently two-component population is real. Assuming the two models
were equally likely before taking the data into account, we find that
the evidence favours the two-component population model by a factor of
$\sim3\times10^7$. \new{Several possible explanations for this result
are explored, namely the evaporation of low-mass stars, a hierarchical
merger, extant remnants of two initially segregated populations, and
multiple star formation epochs.} We find the most compelling
explanation for the two-component velocity distribution is that
\new{47 Tuc formed as two separate populations arising from the same
proto-cluster cloud which merged} $\lesssim7.3\pm1.5$\,Gyr ago.  This
may also explain the extreme rotation, low mass-to-light ratio and
mixed stellar populations of this cluster.
\end{abstract}

\keywords{globular clusters: individual (47 Tucanae)}

\section{Introduction}

As one of the closest and most massive Galactic globular clusters
(GCs), \objectname[47 Tuc]{47 Tucanae} (47 Tuc) is a test-bed for
Galaxy formation models \cite[][and references therein]{Salaris07},
distance measurement techniques \cite[e.g.][]{Bono08} and metallicity
calibrations \cite[e.g.][]{McWilliam08,Lane09b}. \new{This close
examination, however, has left several unresolved conundrums. While
not unique in this respect \cite[see][for a review of elemental
abundance variations in GCs]{Gratton04}, 47 Tuc has a bimodal
distribution of carbon and nitrogen line strengths
\cite[e.g.][]{Harbeck03}. Furthermore, 47 Tuc has a complex stellar
population, exhibiting, for example, multiple sub-giant branches
\cite[e.g.][]{Anderson09}. Although multiple Red Giant and Horizontal
Branches can also be found in other GCs, and may be due to chemical
anomalies \cite[e.g.][]{Ferraro09,Lee09}, 47 Tuc is particularly
unusual in many respects. It has an extreme rotational velocity
\cite[a property it shares only with M22 and $\omega$ Centauri,
e.g.][]{Merritt97,Anderson03,Lane09b} and an apparently unique rise in
its velocity dispersion profile at large radii \cite[][]{Lane09b}.}

Furthermore, the mass-to-light ratio (M/L$_{\rm V}$) of 47 Tuc is very
low for its mass \cite[][]{Lane10}, that is, it does not obey the
mass-M/L$_{\rm V}$ relation described by \cite{Rejkuba07}. Note that
this mass-M/L$_{\rm V}$ relation is not due to the presence of dark
matter but because of dynamical effects
\cite[][]{Kruijssen08}. \new{Explanations for these unusual properties
may be intimately linked to its evolutionary history. In this Letter
we describe and analyse various explanations for the rise in velocity
dispersion in the outer regions of 47 Tuc initially reported by
\cite{Lane09b}.}

\cite{Lane09b} provided a complete description of the data acquisition
and reduction, radial velocity uncertainty estimates, the membership
selection process, and statistical analysis of cluster membership for
all data presented in this Letter.

\section{Plummer Model Fits}\label{Plummerfits}

The \cite{Plummer11} model \cite[see also][]{Dejonghe87} predicts that
the isotropic, projected velocity dispersion $\sigma$ falls off with
radius $r$ as:
\begin{equation}
\sigma(r; \left\{\sigma_0, r_0\right\}) = \frac{\sigma_0}{\left(1 + \left(\frac{r}{r_0}\right)\right)^{1/4}},
\end{equation}
where $\sigma_0$ is the central velocity dispersion and $r_0$ is the
scale radius of the cluster. We now describe how we fitted Plummer
profiles to the 47 Tuc radial velocity data to infer the values of the
parameters $\sigma_0$ and $r_0$, and also to evaluate the overall
appropriateness of the Plummer hypothesis ($H_1$) by comparing it to a
more complex double Plummer model ($H_2$).

The mechanism for carrying out this comparison is Bayesian model
selection \citep{sivia}. Suppose we have two (or more) competing
hypotheses, $H_1$ and $H_2$, with each possibly containing different
parameters $\theta_1$ and $\theta_2$. We wish to judge the
plausibility of these two hypotheses in the light of some data $D$,
and some prior information. Bayes' rule provides the means to update
our plausibilities of these two models, to take into account the data
$D$:
\begin{eqnarray}\label{bayes}
\nonumber\frac{P(H_2|D)}{P(H_1|D)} =
\frac{P(H_2)}{P(H_1)}\frac{P(D|H_2)}{P(D|H_1)} \\ =
\frac{P(H_2)}{P(H_1)}\times\frac{\int_{\theta_1}
p(\theta_1|H_1)p(D|\theta_1, H_1)\,d\theta_1}{\int_{\theta_2}
p(\theta_2|H_2)p(D|\theta_2, H_2)\,d\theta_2}.
\end{eqnarray}
Thus, the ratio of the posterior probabilities for the two models
depends on the ratio of the prior probabilities and the ratio of the
{\it evidence} values. The latter measure how well the models predict
the observed data, not just at the best-fit values of the parameters,
but averaged over all plausible values of the parameters. It should be
noted that we rely solely on the velocity information for our Plummer
model fits. Taking the stellar density as a function of radius into
account would be useful in further constraining the models. However,
the Plummer model fits by Lane et al. (in prep.) based exclusively on
velocity information are also good fits to the surface brightness
profiles of the four GCs analysed in that study.

\subsection{Single Plummer Model}

The data are a vector of radial velocity measurements $\mathbf{v} =
\{v_1, v_2, ..., v_N\}$ of $N$ stars, along with the corresponding
distances from the cluster centre $\mathbf{r} = \{r_1, r_2, ...,
r_N\}$ and observational uncertainties on the velocities
$\mathbf{\sigma_{\textnormal{obs}}} = \{\sigma_{\textnormal{obs}, 1},
..., \sigma_{\textnormal{obs},N}\}$. We will consider $\mathbf{v}$ to
be the data, whereas $\mathbf{r}$ and
$\mathbf{\sigma_{\textnormal{obs}}}$ are considered part of the prior
information. In this case the probability distribution for the data
given the parameters is the product of independent Gaussians, whose
standard deviations vary with radius:
\begin{align}
\nonumber p(\mathbf{v} | \mu, \sigma_0, &r_0)=\\
\prod_{i=1}^N &\left(\frac{1}{\sqrt{2\pi\sigma_i^2}}
\exp\left(-\frac{1}{2}\left(\frac{v_i -
  \mu}{\sigma_i}\right)^2\right)\right),
\end{align}
where $\mu$ is the systemic velocity of the cluster. The standard
deviation $\sigma_i$ for each data point is given by a combination of
the standard deviation predicted by the Plummer model, and the
observational uncertainty:
\begin{equation}
\sigma_i = \sqrt{\sigma(r_i; \left\{\sigma_0, r_0\right\})^2 +
  \sigma_{\textnormal{obs}, i}^2}
\end{equation}
To carry out Bayesian Inference, prior distributions for the
parameters must also be defined. We assigned a uniform prior for $\mu$
(between $-30$ and 30 km s$^{-1}$). For $\sigma_0$, we assigned
Jeffreys' scale-invariant prior $p(\sigma_0) \propto 1/\sigma_0$ for
$\sigma_0$ in the range $0.1-100\textnormal{ km s}^{-1}$. Finally,
$r_0$ was assigned the Jeffreys' prior $p(r_0) \propto 1/r_0$ for
$r_0$ in the range $0.2-220$~pc. These three prior distributions were
all chosen to be independent and to cover the approximate range of
values that we expect the parameters to take.

\subsection{Double Plummer Model}\label{doublePlummer}

The double Plummer model is a simple extension of the Plummer
model. The stars are hypothesised to come from two distinct
populations, each having its own Plummer profile parameters (but with
a common systemic velocity $\mu$). Thus, at any radius $r$ from the
cluster centre, we model the velocity distribution as a mixture
(weighted sum) of two Gaussians. From previous work \cite[][]{Lane09b}
we also expect the inner regions of the cluster to be well fitted by a
single Plummer profile, so the weight of the second population of
stars should become more significant at larger radii.

Thus, instead of having a single $\sigma_0$ parameter and a single
$r_0$ parameter, there are now two of each. The probability
distribution for the data given the parameters is then the weighted
sum of two Gaussians:
\begin{align}
\nonumber p(\mathbf{v} | \mu&, \{\sigma_0\}, \{r_0\}, w(r)) =\\
&\prod_{i=1}^N \left(\frac{w(r_i)}{\sqrt{2\pi\sigma_{i, 1}^2}} \exp\left(-\frac{1}{2}\left(\frac{v_i - \mu}{\sigma_{i, 1}}\right)^2\right)\right. \nonumber\\
&+ \left.\frac{1-w(r_i)}{\sqrt{2\pi\sigma_{i, 2}^2}} \exp\left(-\frac{1}{2}\left(\frac{v_i - \mu}{\sigma_{i, 2}}\right)^2\right)\right).
\end{align}
Here, $w(r)$ is a weight function that determines the relative
strength of one Plummer profile with respect to the other, as a
function of radius. We expect one component to dominate at smaller
radii, and to eventually fade away as the second component becomes
dominant. Hence, we parameterise the function $w(r)$ as:
\begin{eqnarray}
w(r) = \frac{\exp\left(u(r)\right)}{1 + \exp\left(u(r)\right)}
\end{eqnarray}
where
\begin{eqnarray}
u(r) = u_{\alpha} + \frac{r - r_{\textnormal{min}}}{r_{\textnormal{max}} - r_{\textnormal{min}}}\left(u_{\beta} - u_{\alpha}\right).
\end{eqnarray}
That is, the log of the relative weight between one Plummer component
and the other increases linearly over the range of radii spanned by
the data, starting at $u_{\alpha}$ and ending at a value
$u_{\beta}$. Parameterising $w$ via $u$ makes it easier to enforce the
condition that $w(r)$ must be in the range $0-1$ for all $r$. The
prior distributions for $u_{\alpha}$ and $u_{\beta}$ were chosen to be
Gaussian with mean zero and standard deviation 3. This implies that
$w(r)$ will probably lie between 0.05 and 0.95, with a small but not
negligible chance of extending lower than 0.001 or above 0.999.

The standard deviations of the two Gaussians at each data point are
given by a combination of that predicted by the Plummer model and the
observational uncertainty:
\begin{eqnarray}
\sigma_{i, 1} &=& \sqrt{\sigma(r_i; \{\sigma_0, r_0\}_1)^2 + \sigma_{\textnormal{obs}, i}^2} \\
\sigma_{i, 2} &=& \sqrt{\sigma(r_i; \{\sigma_0, r_0\}_2)^2 + \sigma_{\textnormal{obs}, i}^2}
\end{eqnarray}
The priors for all the parameters $\mu$, $\{\sigma_0, r_0\}$ were
chosen to be the same as in the single Plummer case.

\section{Results}

An obvious rise in the velocity dispersion of 47 Tuc was described by
\citet[][their Figure 11]{Lane09b} at approximately half the tidal
radius ($\sim28$\,pc).  The tidal radius is $\sim56$\,pc
\cite[][]{Harris96}. To confirm the reality of this rise, several
tests were performed, including resizing the bins and shifting the bin
centres, as described by \cite{Lane09b}. No difference in the overall
shape of the dispersion profile was found during any tests.

We used a variant \new{\cite[][]{Brewer09}} of Nested Sampling
\citep{skilling} to sample the posterior distributions for the
parameters, and to calculate the evidence values for the single and
double Plummer models. The results are listed in
Table~\ref{results_table}. The double Plummer model is favoured by a
factor $\sim 3 \times 10^7$, and consists of a dominant Plummer
profile that fits the inner parts of the radial velocity data
(Figure~\ref{results_table}), and a second, wider profile that models
the stars at large radius.

\begin{figure*}
\begin{centering}
\includegraphics[scale=0.8]{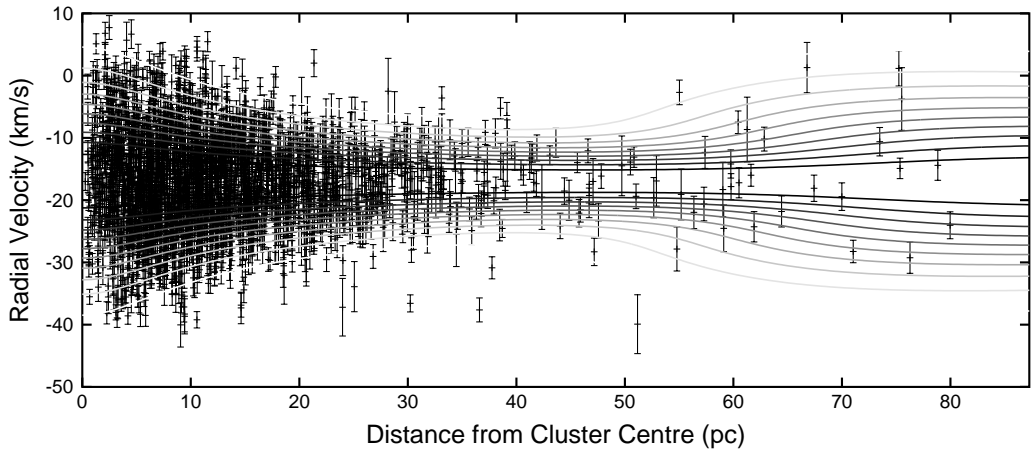}
\caption{The radial velocities of stars in 47 Tuc together with the
best fit double Plummer model for the velocity distribution as a function
of radius. The inner part of the cluster is well modelled by the
dominant Plummer profile, while at larger radii, the second plummer
profile dominates. The radius at which the two profiles have equal
weight is 55 pc.\label{results_figure1}}
\end{centering}
\end{figure*}

\begin{figure*}
\begin{centering}
\includegraphics[scale=0.8]{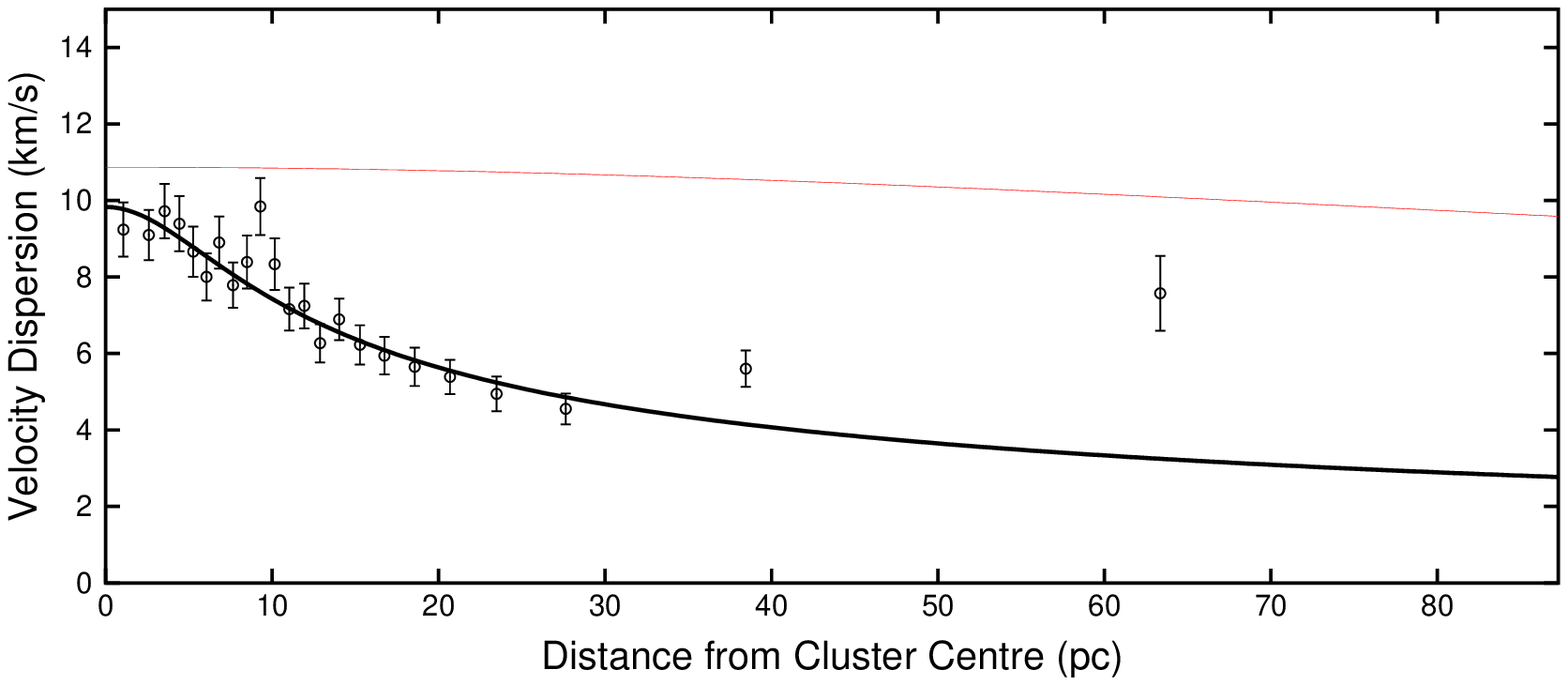}
\caption{Binned velocity dispersion as a function of radius
\citep[from][]{Lane09b}, with the radial velocity profiles of the two
stellar populations from the best fit double Plummer model. The
Plummer profile that dominates at small radii is shown as the thick
black curve, while the thin curve shows the Plummer profile for the
stellar population that dominates at large
radii.\label{results_figure2}}
\end{centering}
\end{figure*}

As a test of the veracity of our model we altered the model so that
$w$($r$) was linear (with endpoints in the range $0-1$ and with a
uniform prior) rather than $u$($r$) being linear. This had the effect
of reducing the log evidence to $\approx-7748$, so in this case the
double Plummer model is favoured by a factor $\sim 10^5$.  This
best-fit model has a more subtle increase in width at large radii,
when compared with Figure \ref{results_figure1}. Presumably this is
because $w$($r$) being linear prevents more rapid fade-outs.
Correspondingly, the Plummer profile represented by the thin curve in
Figure \ref{results_figure2} was shifted slightly lower. Note that the
model in Figure~\ref{results_figure1} is narrower than the spread of
the data, because the spread also arises partly from observational
errors. The most extreme points at large radii are likely to be those
for which the intrinsic velocity dispersion is large {\it and} the
observational errors have pushed the points further away from the
mean. In Figure~\ref{results_figure2}, the Plummer profiles of the two
population components are shown. The inner component is a good fit to
the binned velocity dispersions by \cite{Lane09b}. \new{We now discuss
possible explanations for this two-component population, and calculate
an upper limit on when the second component was introduced.}

\begin{table}
\begin{center}
\caption{Inferred parameter values for the single Plummer and the
double Plummer fits to the 47 Tuc data. The values quoted are the
posterior mean $\pm$ the posterior standard deviation, when the
marginal posterior distributions were sufficiently symmetric for this
to be a reasonable summary. For the few parameters with asymmetric
posterior distributions, we instead give the symmetric 68\% credible
interval. The evidence values imply that if the two models were
equally likely before taking into account the data, the data makes the
double Plummer model more likely by a factor of $e^{17.3} \approx 3
\times 10^7$. For the double Plummer model, the first value listed is
for the component that dominates at $r=0$.}\label{results_table}
\begin{tabular}{lc}
Parameter & Value\\
\hline
{\bf Single Plummer Profile}\\
\hline
$\mu$ & -16.87 $\pm$ 0.17 km s$^{-1}$\\
$\sigma_0$ & 9.37 $\pm$ 0.32 km s$^{-1}$\\
$r_0$ & 9.27 $\pm$ 0.98 pc \\
log(evidence) & -7759.5 \\
\hline
{\bf Double Plummer Profile}\\
\hline
$\mu$ & -16.94 $\pm$ 0.12 km s$^{-1}$\\
$\sigma_0$ & 9.93 $\pm$ 0.43 km s$^{-1}$\\
& [5.70, 13.51] $\pm$  km s$^{-1}$\\
$r_0$ & 6.76 $\pm$ 0.94 pc\\
& [53.6, 4560] pc\\
$u_{\alpha}$ & 6.30 $\pm$ 1.30 \\
$u_{\beta}$& -2.30 $\pm$ 1.12\\
log(evidence) & -7742.2
\end{tabular}
\end{center}
\end{table}

\subsection{Evaporation}\label{evaporation}

\cite{Drukier07} carried out $N$-body simulations of GCs through
core-collapse and into post-core-collapse. They showed that the
evaporation of low-mass stars due to two-body interactions during
these phases alters the velocity dispersion profiles in predictable
ways.  At approximately half the tidal radius ($r_t/2$) the velocity
dispersion reaches a minimum of $\sim40$\% of the central dispersion,
then rises to $\sim60$\% of the central level at $r_t$.  These
criteria are certainly met within 47 Tuc \cite[again, see Figure 11
of][]{Lane09b}. \new{Furthermore, \cite{Lane09b} conclude that the
rise in velocity dispersion could be explained by evaporation due to
the core collapsing, and the Fokker-Planck models by \cite{Behler03}
show that 47 Tuc is nearing core-collapse.}

This appears to be reasonable evidence that 47 Tuc is
evaporating. However, based on the conclusions drawn by
\cite{Drukier07} and \cite{Lane10}, it is unclear how much Galactic
tidal fields affect the outer regions of Galactic globular clusters,
and what effect this has on the external velocity dispersion
profile. \new{While our best-fit double Plummer model matches the
overall form of the trend shown by \cite{Drukier07} reasonably well
(see Figure \ref{results_figure1}), this scenario does not explain its
multiple stellar populations \cite[although these might be explained
by chemical anomalies, e.g.][and references therein]{Piotto07}, nor
its low M/L$_{\rm V}$ in comparison with its mass \cite[see Figure 6
of][]{Lane10}, or its extreme rotation. In addition, the extra-tidal
stars are spread uniformly across all regions of the colour magnitude
diagram \cite[see Figure 3 of][]{Lane09b}, which is inconsistent with
the the two-component population being a consequence of evaporating
low-mass stars. We cannot, however, completely discount evaporation
without detailed chemical abundance information.}

\subsection{Merger}\label{merger}

\new{Another scenario, which appears to explain most of the unusual
properties of 47 Tuc, is that it has undergone a merger in its past
\cite[note that this is not the first evidence for such a merger
within Galactic GCs, see][]{Ferraro02}. Several observed quantities
can be explained by this hypothesis: (1) the bimodality of the carbon
and nitrogen line strengths, (2) the mixed stellar populations, (3)
the large rotational velocity, (4) the low M/L$_{\rm V}$ compared with
total mass and (5) the increase in velocity dispersion in the
outskirts of the cluster.}

\new{In addition to our evidence for two kinematically distinct
stellar populations,} \cite{Anderson09} showed that 47 Tuc also has
two distinct sub-giant branches, one of which is much broader than the
other, as well as a broad main sequence. The authors determined that
this broadening may be due to a combination of metallicities and ages,
and it is known from previous studies \cite[e.g.][]{Harbeck03} that 47
Tuc has a \new{bimodality in its carbon and nitrogen line
strengths. While this bimodality is not unique, a merger could explain
its origin. Therefore, another possible scenario, which explains many
of the properties of 47 Tuc, is a past merger event.}

\new{While it might seem unlikely that this is the remnant of a
merger, extant kinematic signatures of subgalactic scale hierarchical
merging do exist \cite[e.g. within $\omega$ Centauri;][]{Ferraro02},
and there have been hints that the distinct photometric populations in
47 Tuc might be remnants of a past merger \cite[e.g.][]{Anderson09}. A
possible explanation for this merger hypothesis is given in Section
\ref{primordial}.}

\new{\subsection{Initial Formation}\label{primordial}

The two components may have formed at the same epoch, and the distinct
kinematic populations are, therefore, extant remnants from the
formation of the cluster itself. If GCs form from a single cloud
\cite[see][for a discussion of GCs as simple stellar
populations]{Kalirai09}, it is possible for the proto-cluster cloud to
initially contain two overdensities undergoing star formation
independently at almost the same time. In this case, the two
proto-clusters, which would inevitably be in mutual orbit due to the
initially bound nature of the proto-cluster cloud, eventually
coalesced through the loss of angular momentum due to dynamical
friction.

Note that this scenario is similar to the capture of a satellite
described by \cite{Ferraro02} to explain the merger hypothesis for
$\omega$ Centauri, and explains the two kinematic and photometric
populations, and the high rotation rate assuming the proto-cluster
cloud initially had a large angular momentum. It might also explain
the low ${\rm M/L_{V}}$ of this cluster.

Interestingly, \cite{Vesperini09} show that clusters with initially
segregated masses evolve more slowly than non-segregated clusters,
having looser cores and reaching core-collapse much later. Because the
core of 47 Tuc is highly concentrated and near core-collapse
\cite[e.g.][]{Behler03} it must be very old if it was initially
mass-segregated. 47 Tuc is thought to be 11--14\,Gyr old
\cite[e.g.][]{Gratton03,Kaluzny07}, hence initial mass segregation is
plausible. However, even if 47 Tuc is $\sim11$\,Gyr old the original
populations would be kinematically indistinguishable at the present
epoch (Section \ref{time}) indicating that some other process was the
cause of the two extant populations described in this
Letter. Furthermore, Milky Way GC formation ended about 10.8\,Gyr ago
\cite[e.g.][]{Gratton03}, long before the upper limit for the initial
mixing of the two populations derived in Section \ref{time}.

\subsection{Multiple Star Formation Epochs}\label{SF}

Two star formation epochs in GCs result in the radial separation of
the two populations, with the second generation initially concentrated
in the core \cite[e.g.][and references
therein]{D'Ercole08}. Furthermore, the kinematics of the second
generation are virtually independent of that of the first generation
and the second generation contain chemical anomalies which are
consistent with having arisen in the envelopes of the first generation
\cite[e.g.][]{Decressin07,D'Ercole08}.

This scenario might explain the two kinematic populations and chemical
anomalies of 47 Tuc, however, it is unclear how this would cause the
extreme rotation, or the anomalous ${\rm M/L_{V}}$.}

\subsection{Time-scale for the Initial Mixing of the Second Population}\label{time}

\new{The scenarios described in Sections \ref{merger},
\ref{primordial}, and \ref{SF} require a second population beginning
to mix with an initial population at a particular epoch.}
\cite{Decressin08} performed detailed $N$-body simulations of globular
clusters containing two distinct populations of stars to determine
their dynamical mixing time via two-body relaxation. They concluded
that $\sim2$ relaxation times are required to completely homogenise
the populations, a timescale that is virtually independent of the
number of stars in the cluster. They also concluded that the
information on the stellar orbital angular momenta of the two
populations is lost on a similar timescale. Since we observe two
distinct, extant kinematic populations\new{, an upper limit can be
placed on when the two populations began to mix.}

\cite{Decressin08} showed that the relaxation time ($t_{rh}$) of a GC
decreases by $\sim0.29t_{rh}$ for each consecutive relaxation time,
i.e.  $t_{rh}(i)\sim0.71t_{rh}(i+1)$. Because the two populations are
kinematically distinct at the present epoch, and the current
relaxation time of 47 Tuc is $t_{rh}\approx3.02$\,Gyr
\cite[][]{Harris96}, \new{an upper limit on when the two populations
began to mix} is $7.3\pm1.5$\,Gyr ago. \new{Note that the scenarios
discussed in Sections \ref{merger} and \ref{primordial} can be
temporally assessed in this way only if the merger did not disrupt the
core of the larger component. This is true for minor mergers in blue
compact dwarf galaxies \cite[][]{Sung02}, therefore, if we assume a
merger between objects with a mass ratio of 9:1 \cite[][show the ratio
of the number densities on the two sub-giant branches of 47 Tuc is
$\sim$9:1]{Anderson09}, this is also likely to hold for 47 Tuc.}

\section{Conclusions}

With a Bayesian analysis of the velocity distribution of 47 Tuc, we
conclude that \new{the scenario which best explains the observed
properties of 47 Tuc is that} we are seeing the first kinematic
evidence of a merger in 47 Tuc, which occurred
$\lesssim7.3\pm1.5$\,Gyr ago. \new{Extant kinematic populations from
the merger of formation remnants is a plausible explanation as to the
reason for this merger, assuming the two components evolved separately
and merged $\lesssim7.3\pm1.5$\,Gyr ago.} This scenario \new{could
explain the two-component population,} its extreme rotational
velocity, mixed stellar populations and low M/L$_{\rm V}$ compared
with its mass.

\new{All the other explanations for this two-component population are
less plausible than the merger hypothesis. Evaporation of low mass
stars is unlikely due to the various stellar types that are found
beyond the tidal radius, and this also cannot explain its low
M/L$_{\rm V}$. The possibility of multiple star formation epochs does
not explain the large rotational velocity, nor the low M/L$_{\rm V}$.}

 \new{Detailed chemical abundances and} high resolution $N$-body
simulations of merging globular clusters are now required to further
\new{analyse the} merger scenario. \new{Several observed quantities
need to be} addressed, namely how much angular momentum can be
imparted through a 9:1 merger, what consequence it has on the velocity
dispersion in the outer regions of the cluster over dynamical
timescales, and what effect it would have on the global M/L$_{\rm V}$.

Of course, alternative explanations exist for the observed rise in
dispersion.  For example, if GCs form in a similar fashion to Ultra
Compact Dwarf galaxies, there may be a large quantity of DM in the
outskirts of the cluster as discussed by \cite{Baumgardt08}.  However,
no evidence exists supporting GCs forming in this manner and GCs do
not appear to have significant dark matter components
\cite[e.g.][]{Lane09a,Lane09b,Lane10}.

\acknowledgments

This project has been supported by the University of Sydney, the
Anglo-Australian Observatory, the Australian Research Council, the
Hungarian OTKA grant K76816 and the Lend\"ulet Young Researchers
Program of the Hungarian Academy of Sciences. \new{GyMSz acknowledges
the Bolyai Fellowship of the HAS.} RRL thanks Martine L. Wilson for
everything.

\end{document}